**From Evidence to Design: Developing an AI-Augmented UX Research Point of View for Digital Wellbeing in Emergency and Public Safety Contexts**

Olumuyiwa Ayorinde
School of Computing and Engineering, Bournemouth University Poole, UK,
ayorindeo@bournemouth.ac.uk

Huseyin Dogan
School of Computing and Engineering, Bournemouth University Poole, UK,
hdogan@bournemouth.ac.uk

Festus Adedoyin
School of Computing and Engineering, Bournemouth University Poole, UK,
fadedoyin@bournemouth.ac.uk

Nan Jiang
School of Computing and Engineering, Bournemouth University Poole, UK,
njiang@bournemouth.ac.uk

Emmanuel Oluokun
School of Computing and Engineering, Bournemouth University Poole, UK,
eoluokun@bournemouth.ac.uk

Abiodun Adedeji
School of Computing and Engineering, Bournemouth University Poole, UK,
adedejia@bournemouth.ac.uk

Melike Akca
School of Computing and Engineering, Bournemouth University Poole, UK,
makca@bournemouth.ac.uk

This paper investigates how User Experience Research (UXR) methods can be combined with AI-supported analysis to develop clearer design direction for digital wellbeing interventions targeting Emergency and Public Safety Personnel (EPSP). EPSP work in high-stress, shift-based environments where cognitive fatigue and unpredictable schedules reduce engagement with conventional wellbeing tools. Using the UXR Point-of-View (PoV) framework, this study applied an AI-supported literature analysis process to identify recurring psychological, behavioural, and design patterns. Behaviour Change Techniques and Persuasive Technology principles were integrated throughout interpretation to connect evidence with practical design reasoning. The process resulted in a UXR PoV Pyramid, nine UXR Play Cards, and stakeholder focused PoV narratives. Findings show that effective wellbeing systems for EPSP must minimise cognitive effort, adapt to operational context, and prioritise psychological safety. The work demonstrates how AI can assist large-scale evidence interpretation while human researchers maintain responsibility for contextual judgement and design direction.

CCS CONCEPTS • Human Computer Interaction• User Experience Research• Artificial intelligence • Assistive technologies

Additional Keywords and Phrases: Digital health and wellbeing; Emergency and public safety personnel (EPSP); User Expereince Research (UXR); UXR Point of View (PoV), AI-augmented analytical approach, shift workers, Human-centered AI

## 1 INTRODUCTION

Unlike traditional 9–5 work settings, individuals employed in high-stress occupations, particularly those working in shift-based systems, such as emergency and public safety personnel (EPSP) (e.g., police officers, ambulance personnel, and firefighters), operate within environments defined by strict operational demands and often unpredictable schedules. These work environments often involve irregular hours, last-minute changes, extended shifts, and continuous exposure to high-pressure situations [1]. Research has shown that shift workers experience elevated levels of occupational stress, increased injury risk, and reduced quality of life [2,3]. Shift-related stress is also associated with poor sleep, fatigue, mental health challenges, cardiovascular risks, reduced work–life balance, and long-term physical and cognitive health issues [4]. These conditions often make sustained interaction with digital wellbeing tools difficult [1, 4].

Digital health technologies (DHTs) have been increasingly positioned as a means of promoting healthier behaviours and supporting overall wellbeing [5]. These technologies are widely used to support, monitor, and manage health and wellbeing outcomes [5,6], and has the potential to support underserved populations [7], such as emergency and public safety personnel (EPSP). However, many existing digital health tools are designed around users with stable routines and conventional work schedules, making them poorly suited for shift workers [1]. EPSP roles are characterised by unpredictable operational demands, exposure to trauma, and fluctuating workloads, which can negatively affect engagement with wellbeing technologies [4,8]. As a result, usability, sustained use, and long-term adoption remain limited.

Mehra et al. [9] reported that officers recommended shift-management features to track wellbeing over time. The lack of schedule awareness in many mobile health applications remains a major gap, contributing to poor suitability for emergency service personnel [10]. Although digital wellbeing tools are widely available, their effectiveness often remains limited because design decisions fail to account for contextual constraints such as unpredictable schedules, cognitive fatigue, and variable energy levels [11,12].

Behaviour changes technologies increasingly inform digital health design. Behaviour Change Techniques (BCTs), such as goal setting, self-monitoring, and social support, are commonly used to encourage positive wellbeing outcomes [13], while persuasive technologies use personalised prompts and tailored suggestions to support engagement [14]. Digital interventions are particularly suited to behaviour change due to their scalability and accessibility [15]. These approaches have shown promise across domains including physical activity, smoking cessation, and healthy eating [16], targeting outcomes such as motivation, self-efficacy, and adherence. However, behaviour change is shaped by psychological, social, and environmental factors [17], and existing frameworks offer limited guidance for translating complex evidence into practical design decisions.

UX research often generates large volumes of data, yet teams struggle to translate findings into clear, actionable design direction. Without structured approaches, evidence can be interpreted inconsistently, reducing impact and alignment. The challenge is therefore not a lack of theory, but a lack of practical methods for converting evidence into context-sensitive design guidance.

The UXR Point-of-View (PoV) framework addresses this gap by providing a structured pathway from research evidence to strategic design direction [18]. Organised as a four-level pyramid (foundation, data, insight, and PoV), it supports shared understanding across multidisciplinary teams, while the UXR PoV Playbook uses plays and play cards to translate complex findings into actionable outcomes. The framework has been applied across multiple domains, demonstrating its adaptability beyond single contexts, and supporting its use in varied design and research environments [19,20].

In this study, Generative AI was used as a support tool within the UXR PoV framework to help organise literature evidence, identify recurring patterns, and assist early hypothesis development. Rather than replacing researchers, Generative AI acted as a collaborator that supported evidence organisation, pattern identification, and consistent interpretation, while human researchers retained responsibility for contextual judgement and decision-making [21].

The process followed the four-stage structure introduced in the Developing an AI-Powered UX Research Point of View (PoV) framework [21]:

1. Leveraging GenAI and the UXR PoV approach
2. Establishing a foundational plan and roadmap
3. Insight generation and UXR Play Card development
4. PoV narrative construction and stakeholder communication

Building on this approach, we integrated behaviour change and persuasive technology frameworks to develop design recommendations for digital wellbeing interventions targeting Emergency and Public Safety Personnel (EPSP). The goal is to demonstrate how AI-augmented UXR PoV approach can help researchers articulate clearer, context-sensitive perspectives suited to high-stress occupational environments.

## METHOD

### 2.1 Study Design Overview

This study applied an AI-augmented systematic literature review (SLR) combined with User Experience Research (UXR) synthesis using the UXR Point-of-View (PoV) framework. The method integrates PRISMA-based evidence synthesis, organised through Excel dataset, with Generative AI (GenAI)–assisted reasoning to translate wellbeing research relating to Emergency and Public Safety Personnel (EPSP) into actionable UX Points of View for digital wellbeing interventions. The methodological process followed the four-stage structure of the *Developing an AI-Powered UX Research Point of View (PoV) workshop* [21]: (1) leveraging GenAI and the UXR PoV approach; (2) establishing a foundational plan and stakeholder roadmap; (3) insight generation and UXR Play Card development; and (4) PoV narrative construction and stakeholder communication.

A systematic literature search was conducted across relevant academic databases, including PubMed, and Scopus, covering publications related to digital health and wellbeing interventions for Emergency and Public Safety Personnel (EPSP). Search terms combined wellbeing, shift work, emergency services, and digital interventions (e.g., mobile applications, digital wellbeing tools). Studies were screened according to predefined inclusion and exclusion criteria. All included studies were coded in Microsoft Excel using a PRISMA-aligned structure, capturing bibliographic information, participant characteristics, intervention type, wellbeing outcomes, and design-relevant findings. To support AI-assisted analysis, the following materials were uploaded to the Gen-AI platform:

1. the PRISMA-organised Excel dataset,
2. the UXR PoV framework papers [20, 21] and related publications,
3. UXR previous publication and workshop materials,
4. example Play Cards and UXR PoV Playbook website.

### 2.2 Procedure and Prompts

**Stage 1: AI- Assisted Analysis and the UXR PoV Framework**

The goal was to extract recurring design-relevant themes and identify user needs, pain points, and contextual barriers from the systematic review data. GenAI (Chat GPT 5.2) supported the analysis by identifying recurring behavioural, contextual, and experiential patterns related to EPSP wellbeing interventions. Particular attention was given to shift work constraints, cognitive load, engagement barriers, and trust concerns. Prompts included:

- *"Analyse this SLR Excel dataset and UXR design PoV framework. Identify user challenges, barriers across digital health and wellbeing intervention studies for emergency and public safety personnel."*
- *"Cluster these insights into UX research themes that align with the UXR Point of View Framework."*
- *"Analyse the provided qualitative and quantitative user data to identify recurring patterns, contextual factors, and comprehensive unmet user needs."*
- *Based on these synthesised themes, generate potential hypotheses and design opportunities relevant to health and wellbeing interventions for emergency and public safety personnel*

This stage helped surface recurring user challenges, emerging themes, and early hypotheses grounded in the systematic review data.

**Stage 2: Establishing a Foundational Plan and Stakeholder Roadmap**

A shared conceptual foundation linking user needs, behavioural theory, and stakeholder goals was built here. GenAI was used to help translate extracted themes into a roadmap for intervention design. Prompts included:

- *"Based on the systematic review insights, explain how emergency and public safety personnel individuals typically manage health and intervention, and what digital interventions can realistically support their* *health and wellbeing needs tailored to their shift work pattern without increasing cognitive overload with better adoption and engagement"*
- *"Using the extracted themes, help me formulate clear user experience and health and wellbeing goals for a digital intervention."*
- *"Who are the primary and secondary stakeholders in a project developing digital health and wellbeing tools for emergency and public safety personnel? Describe their roles, motivations, and data needs"*
- *"Create a step-by-step project plan linking research goal, user needs, and stakeholder engagement following a mixed-method approach (quantitative, qualitative, AI synthesis)"*
- *"Identify which constructs best explain health and wellbeing challenges of emergency and public safety personnel, using BCT and persuasive tech frameworks"*

Insights were grounded in stakeholder context by mapping user needs, organisational constraints, and data considerations. Behaviour Change Techniques and Persuasive Technology principles were used as interpretive lenses to explain motivational and behavioural mechanisms.

**Stage 3: Insight Generation and Play Card Development**

This stage focused on transforming evidence and theory into design insights using the UXR PoV building blocks: Foundation, Data, Insight Generation, and PoV operationalisation. GenAI was used iteratively to connect empirical findings with behavioural interpretation and UX implications. Behavioural models were integrated directly into analytical workflow to explain why specific design mechanisms may support or hinder engagement. Prompts included:

- *"Using BCT, Persuasive tech, develop hierarchical design principles for health and wellbeing apps targeting emergency and public safety users."*
- *"Interpret these patterns through a UXR lens, explain why certain feedback work or fail, and make them actionable for design."*
- *"Turn these validated insights into concise UX Points of View, one sentence per insight, explaining the user's emotional need and recommended design approach"*
- *"Develop the UXR Play Card Library."*
- *"Translate the UXR Point of View into a reusable Play Card by defining its goal, when to apply it, why it matters, and how it can be implemented"*

**Stage 4: Designing PoV Narratives and Stakeholder Communication**

Play Card insights were consolidated into stakeholder-facing PoV narratives using the UXR PoV framework to support communication clarity and tone across research, design, and decision-making contexts. Prompts included:
- *"Populate a UXR POV using the UXR POV Template structure."*
- *"Condense this template into an executive-ready POV narrative."*
- *"Reframe the validated POV for a design audience."*
- *"Emphasise practical design implications while maintaining accountability and human-in-the-loop positioning."*

## 3 RESULTS

This study produced a structured set of findings demonstrating how Generative AI (GenAI) can support the development of a User Experience Research (UXR) Point of View (PoV) for digital wellbeing interventions targeting Emergency and Public Safety Personnel (EPSP). Results are organised according to the four-stage AI-augmented UXR PoV process: (1) leveraging GenAI for evidence analysis and early insight generation, (2) establishing a foundational stakeholder-informed roadmap, (3) translating insights into a UXR Play Card library, and (4) crafting stakeholder-aligned UX Point-of-View narratives. Across these stages, behavioural and persuasive design principles were used to interpret emerging patterns and support translation into actionable design direction.

The results present a progression from evidence interpretation toward design articulation, reflecting the layered structure of the UXR PoV framework (Figure 1). Rather than replacing analysis, GenAI supported pattern identification and structuring of findings, while researchers-maintained responsibility for contextual interpretation and decision-making within high-stress EPSP environments.

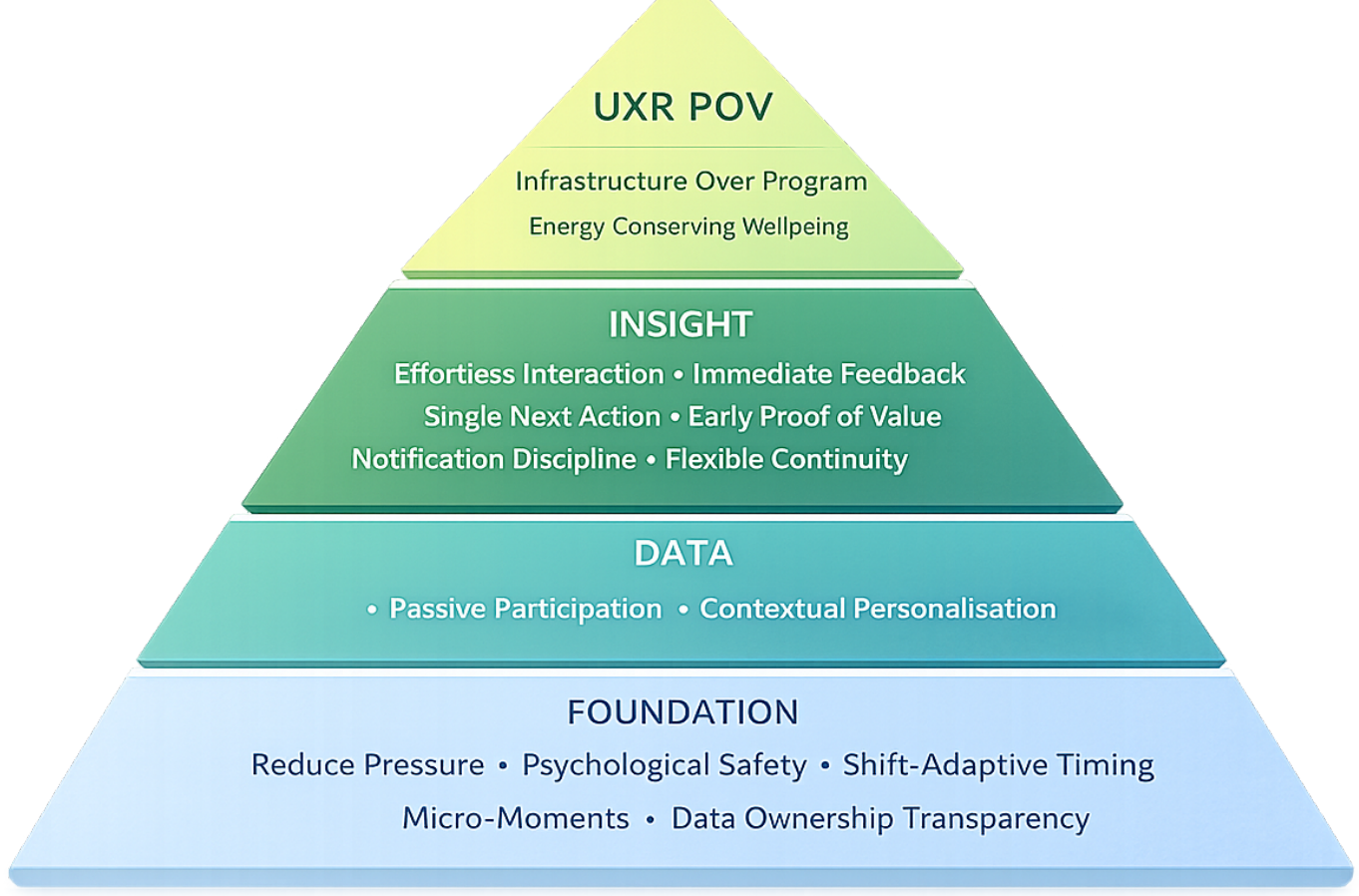

Figure 1. UXR Point of View (PoV) Pyramid Framework.
*(The UXR Point of View (PoV) Pyramid adapted for Emergency and Public Safety Personnel (EPSP) digital wellbeing intervention)*

**3.1 Leverage GenAI and the UXR PoV Framework**

In the first stage, Generative AI (ChatGPT 5.2) was used to support evidence analysis and early insight generation from the literature examining digital wellbeing interventions for Emergency and Public Safety Personnel (EPSP). Analysis revealed recurring patterns across psychological, behavioural, and design-related dimensions.

The psychological layer captured themes including cognitive depletion, emotional strain, stigma concerns, and trust uncertainty. The behavioural layer highlighted inconsistent engagement, notification avoidance, short interaction windows, and resistance toward interventions perceived as additional work. The design layer revealed recurring needs for low-effort interaction, contextual timing, flexible participation, and transparent data practices.

Through iterative prompting and researcher-led validation, these themes were translated into a set of design-oriented hypotheses linking interaction design characteristics to engagement outcomes. The hypotheses generated in this stage were:

- **H1:** Low-effort interaction increases engagement when users experience cognitive depletion and limited time or attention
- **H2:** Shift-aware intervention timing improves perceived relevance and reduces notification dismissal compared to fixed-time interventions.
- **H3:** Psychological safety and trust-centred design increase onboarding completion and sustained participation.
- **H4:** Interventions designed for short interaction windows increase engagement within time-constrained operational environments.
- **H5:** Reducing manual tracking through passive or lightweight participation increases data contribution and sustained use.
- **H6:** Visible short-term feedback and micro-wins improve re-engagement and perceived usefulness.
- **H7:** Clear communication of data visibility and ownership increases trust and willingness to engage.
- **H8:** Wellbeing tools framed as always-available infrastructure support opportunistic engagement more effectively than structured programmes.
- **H9:** Systems that allow intermittent use without penalties support long-term engagement in irregular work environments.

These hypotheses served as foundational knowledge artefacts guiding subsequent stakeholder analysis and design translation.

**3.2 Establish a Foundational Plan and Stakeholder Roadmap**

In the second stage, GenAI-supported synthesis was used to establish a shared conceptual foundation linking user needs, organisational constraints, and design goals. The analysis identified a diverse stakeholder ecosystem including EPSP end users, team leaders, clinicians and wellbeing providers, UX researchers and designers, technology developers, and organisational decision-makers.

Two recurring tensions emerged. The first concerned privacy versus organisational oversight, where users expressed concern about how wellbeing data might be interpreted or accessed. The second involved the mismatch between operational realities and intervention expectations, where structured wellbeing programmes conflicted with unpredictable schedules and fluctuating energy levels.

This stakeholder mapping (Table 1) clarified how design solutions must balance trust, usability, and organisational needs while remaining adaptable to operational contexts. The roadmap developed at this stage informed the development of practical design artefacts in the next phase.

| Stakeholder | Role | Need | Challenge |
|---|---|---|---|
| EPSP Users (Police, Fire, Ambulance) | End-Users | Wellbeing support compatible with unpredictable shifts and cognitive load | Privacy concerns and fear of organisational monitoring; limited time and fluctuating energy |
| Team Leaders / Managers | Operational Supervisors | Support workforce wellbeing without disrupting operational delivery | Balancing oversight responsibilities with employee trust and data sensitivity |
| Clinicians / Wellbeing Providers | Care Facilitators | Ethical, trust-centred intervention pathways | Integrating digital wellbeing signals without compromising autonomy |
| UX Researchers / Designers | Translators | Converting behavioural insights into context-aware design principles | Designing for operational unpredictability and stigma-sensitive engagement |
| Technology Developers | System Builders | Implementable, adaptive system logic | Balancing passive support mechanisms with privacy-preserving architecture |
| Organisational / Policy Stakeholders | Strategic Decision-Makers | Sustainable adoption and measurable wellbeing outcomes | Addressing low engagement and high attrition in structured wellbeing programmes |

*Table 1. Stakeholder Mapping within Gen-AI-Assisted UXR PoV Framework*

### 3.3 Apply GenAI-Enhanced Best Practices

In the third stage, theory-informed hypotheses were translated into a structured set of nine UXR Play Cards. Developed with support from ChatGPT 5.2 and refined through researcher interpretation, the cards operationalise behavioural insights and empirical evidence into practical design mechanisms for UX researchers, designers, and interdisciplinary teams working within Emergency and Public Safety Personnel (EPSP) wellbeing contexts.

Rather than functioning as isolated recommendations, the cards represent core design responses to recurring challenges identified across earlier stages. A primary theme concerned interaction effort and cognitive depletion. Cards such as Effortless Interaction emphasise reducing decisions, simplifying flows, and supporting low-energy states through minimal or guided interaction, reflecting evidence that perceived effort strongly shapes disengagement.

A second theme addressed the temporal realities of EPSP work. Shift-Adaptive Timing (figure 3) and Micro-Moment Support (figure 4) highlight the need for brief, context-aware interactions that align with unpredictable schedules and recovery windows. These cards reinforce the finding that engagement is opportunistic rather than routine, requiring flexible rather than programme-based models.

Trust and psychological safety (figure 2) formed another central cluster. Psychological Safety First and Data Ownership Transparency (Figure 5) translate concerns around stigma, judgement, and data visibility into design strategies centred on neutral framing, clear privacy boundaries, and user control. These mechanisms reflect stakeholder insights showing that perceived safety is a prerequisite for sustained use.

At the system level, Infrastructure Over Program reframes digital wellbeing tools as always-available support embedded within daily workflows rather than structured programmes demanding ongoing commitment. Supporting mechanisms such as Passive Participation, Immediate Reassurance Feedback, and Flexible Continuity further emphasise low-pressure engagement, reduced self-report burden, and visible early value.

Collectively, the nine cards function as boundary objects that connect behavioural reasoning with practical UX decision-making (Table 2). Each card captures a core challenge and translates it into actionable guidance through a consistent structure: a defined issue, associated risks, and evidence-aligned best-practice recommendations. Together, they establish a shared design language that enables interdisciplinary teams to move from research insight toward coherent design direction while remaining grounded in operational realities rather than generic wellbeing assumptions.

| Card | Title | Quote | Issue Type | Best Practice | UXR Skills |
|---|---|---|---|---|---|
| **1** | Effortless Interaction | Design for depleted minds, not ideal conditions. | Cognitive Depletion | Enable one-tap actions; reduce decisions; simplify flows; support low-energy interaction modes. | Cognitive Psychology, Interaction Design, Accessibility Design |
| **2** | Psychological Safety First | Trust precedes engagement. | Stigma & Trust Uncertainty | Use neutral, non-clinical language; apply strength-based framing; clarify privacy boundaries; reinforce user control. | Trust-Centred Design, Ethical UX, Emotional Design |
| **3** | Shift-Aware Timing | Timing determines receptivity. | Contextual Misalignment | Align prompts with shift transitions and recovery windows; reduce notification frequency; support quiet modes. | Temporal UX, Adaptive Systems Design |
| **4** | Micro-Moment Support | Support must fit the moments users actually have. | Operational Time Scarcity | Design interactions under 60 seconds; remove rigid routines; enable optional, lightweight engagement. | Behavioural Design, Experience Strategy |
| **5** | Passive Participation | Measurement should not feel like work. | Self-Report Burden | Use passive sensing where possible; minimise manual tracking; surface early value automatically. | Behavioural Analytics, UX Research Methods |
| **6** | Immediate Reassurance Feedback | Visible benefit sustains engagement. | Delayed Perceived Value | Provide real-time feedback; highlight micro-wins; show small improvements quickly. | Motivation Design, Affective UX |
| **7** | Data Ownership Transparency | Clarity builds trust. | Data Governance & Privacy Concern | Clearly communicate who sees what data; separate personal vs organisational visibility; reinforce privacy messaging. | Ethical UX, Governance Design |
| **8** | Infrastructure Over Program | Wellbeing works best as infrastructure, not a programme. | Programme Attrition | Provide always-available tools; allow ambient use; adapt support to context and role. | Systems Thinking, Service Design |
| **9** | Flexible Continuity | Users should return without penalty. | Inconsistent Engagement Patterns | Allow intermittent use; preserve progress across gaps; remove streak pressure or failure framing. | Experience Design, Longitudinal UX Strategy |

*Table 2. GenAI-derived UXR Play Cards for EPSP wellbeing design*

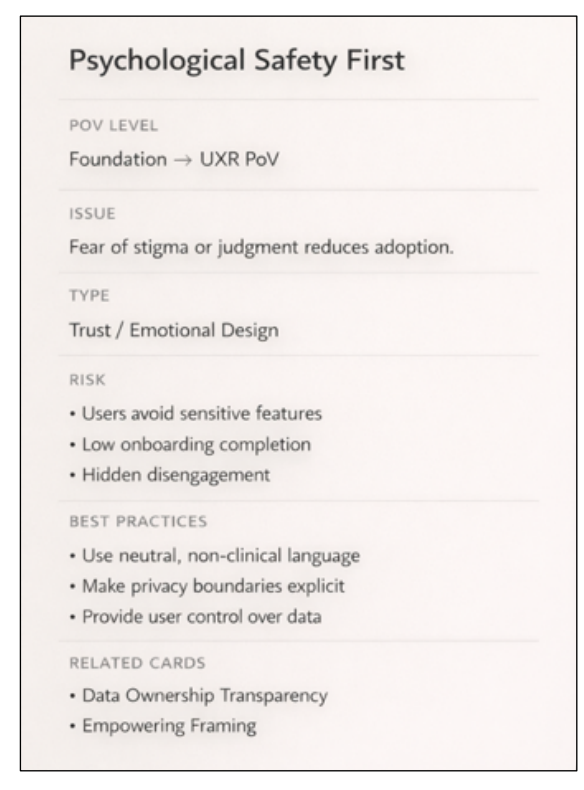

*Figure 2: UXR Play Card 2 Psychological Safety First*

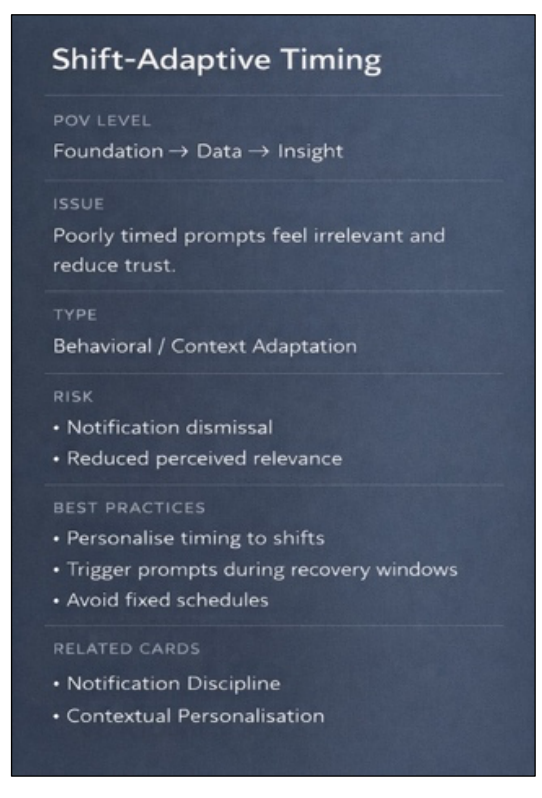

*Figure 3: UXR Play Card 3 Shift-Aware Timing*

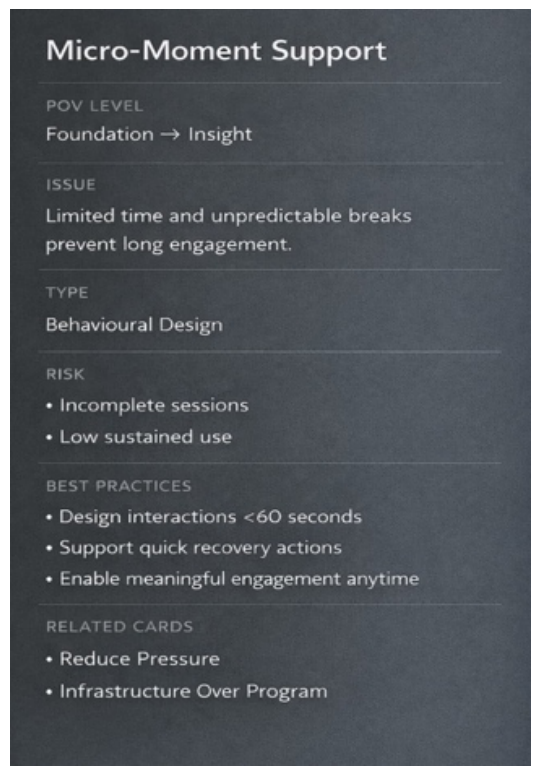

*Figure 4: UXR Play Card 4 Micro-Moment Support*

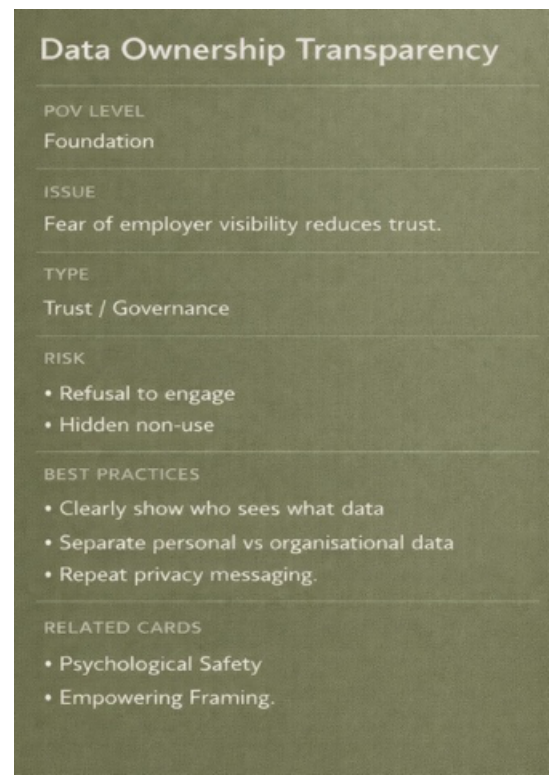

*Figure 5: UXR Play Card 7 Data Ownership Transparency*

### 4.4 Crafting PoV Narratives and Stakeholder Communication

In the fourth stage, insights derived from the UXR Play Cards were translated into stakeholder-adapted Point-of-View (PoV) narratives using the existing UXR PoV narrative template described in the UXR PoV Playbook. The purpose of this stage was to move from design insights toward clear, role-specific communication that supports decision-making across research, design, and implementation contexts.

The PoV narrative template provided a structured format for articulating design direction by linking evidence, behavioural reasoning, and practical implications. Using this structure, GenAI supported iterative phrasing and organisation of narrative content, while researchers ensured alignment with EPSP operational realities and maintained clarity of meaning. This process ensured that the final PoV narratives remained grounded in evidence while being understandable and actionable for different stakeholder groups.

Stakeholder-adapted PoV narratives were developed for key groups identified earlier in the methodological process, including EPSP end users, organisational leaders, clinicians and wellbeing providers, UX researchers, designers, and technology developers. For EPSP users, narratives emphasised low-effort interaction, psychological safety, and contextual adaptability. Organisational stakeholders focused on sustainable adoption without operational disruption, while clinicians prioritised trust-centred and ethically aligned intervention pathways.

Importantly, UX researchers and designers were supported through narratives that translated behavioural findings into clear design reasoning and reusable guidance, improving user experience clarity across development stages. Developers received narratives focused on feasibility, adaptive system behaviour, and low-friction implementation, ensuring that technical decisions aligned with behavioural and contextual insights identified in earlier stages.

Across stakeholder perspectives, a shared PoV emerged: digital wellbeing interventions for EPSP are most effective when they minimise cognitive effort, adapt to operational context, and function as flexible infrastructure rather than structured programmes (Table 3). The stakeholder-adapted narratives therefore served as a communication bridge between research evidence and implementation, supporting alignment across interdisciplinary teams.

| Stakeholder | Core Need | Primary Challenge | PoV Narrative Focus | GenAI Contribution |
|---|---|---|---|---|
| **EPSP Users (Police, Fire, Ambulance)** | Wellbeing support that fits unpredictable schedules | Cognitive fatigue, limited time, stigma concerns | Low-effort interaction, shift-aware support, psychological safety | Theme clustering across studies to identify recurring engagement barriers |
| **Team Leaders / Managers** | Workforce wellbeing without operational disruption | Supporting wellbeing while maintaining trust | Aggregated, non-intrusive wellbeing insights | Identification of organisational tensions between monitoring and privacy |
| **Clinicians / Wellbeing Providers** | Ethical, supportive intervention pathways | Balancing care visibility with autonomy | Trust-centred design | Mapping behavioural signals to support opportunities |
| **UX Researchers / Designers** | Translating evidence into design direction | Moving from data to actionable insight | Play Cards and PoV synthesis as design guidance | Narrative generation and insight organisation |
| **Technology Developers** | Implementable design requirements | Reducing friction while maintaining functionality | Passive interaction, adaptive timing, simplified flows | Pattern identification linking behaviour and design mechanisms |
| **Organisations / Policy Stakeholders** | Sustainable adoption and measurable outcomes | High attrition and low engagement in wellbeing tools | Infrastructure-over-program strategy | Analysis of adoption barriers across evidence sources |

*Table 3. GenAI-Assisted PoV Narrative Development within the UXR PoV Framework*

Together, these stakeholder-adapted PoV narratives demonstrate how structured GenAI support can translate behavioural evidence into actionable, context-aware guidance while maintaining human oversight. While this enhances alignment and clarity across interdisciplinary teams, it also raises important questions about methodological robustness and the limits of AI-assisted synthesis in complex socio-technical systems. The following discussion critically reflects on these strengths and limitations.

## 4 DISCUSSIONS

This study examined how Generative AI (GenAI) can support User Experience Research (UXR) in developing digital wellbeing interventions for Emergency and Public Safety Personnel (EPSP). By combining behavioural and persuasive design principles [13,14,17] with AI-assisted evidence analysis, the work aimed to develop digital wellbeing systems suited to high-stress, shift-based occupational environments [1]. The findings indicate that GenAI can meaningfully support analytical work, but its outputs require careful human interpretation to remain contextually valid.

GenAI proved effective in tasks such as clustering engagement patterns, organising findings across studies, and generating early hypotheses. These capabilities accelerated identification of recurring barriers including cognitive fatigue, timing misalignment, and trust concerns, which informed the development of nine UXR Play Cards. The cards translated behavioural observations into practical design mechanisms, helping move from evidence toward design reasoning. However, AI-generated outputs sometimes failed to fully account for operational realities specific to EPSP contexts, including unpredictable schedules, organisational culture, and workload intensity.

These limitations became clearer when stakeholder perspectives were considered. EPSP users prioritised psychological safety and low-effort interaction, while organisational stakeholders emphasised adoption and measurable outcomes. Designers focused on usability and clarity, whereas developers prioritised feasibility and performance. The resulting tensions highlighted the challenge of defining effective wellbeing support across disciplines. UXR Play Cards functioned as boundary objects that helped align behavioural evidence with practical design decisions. A broader issue relates to AI systems themselves. While models efficiently identified dominant patterns, some outputs reflected generic wellbeing assumptions that conflicted with interaction patterns found in shift-based work. Without human review, such outputs risk reinforcing programme-style engagement models that do not fit operational realities. This reinforces the need for researcher-led interpretation when applying AI within specialised domains.

Accordingly, this study adopted a researcher-led, AI-assisted approach in which GenAI extended analytical capacity while human researchers retained responsibility for contextual judgement. Human-centred design remained essential for translating evidence into interventions that align with real-world workflows and organisational pressures. For UXR PoV practice [20], the implications are clear. GenAI (ChatGPT 5.2) can reduce analytical workload and support early exploration, but responsible use requires iterative validation, stakeholder grounding, and bias-aware review. When these safeguards are absent, AI-generated insights may appear coherent yet fail under real operational conditions.

Methodologically, this work extends the UXR PoV framework by showing how GenAI can support multiple stages of the PoV pyramid, from evidence interpretation and hypothesis development to Play Card creation and stakeholder narrative articulation. The resulting Play Cards demonstrate a practical way to convert behavioural evidence into accessible design artefacts that support collaboration across multidisciplinary teams [21].

## 5 CONCLUSION AND FUTURE WORK

This research presents a GenAI-augmented UXR Point-of-View approach for designing digital wellbeing interventions aligned with the realities of Emergency and Public Safety Personnel. By combining AI-assisted analysis with behavioural and persuasive design frameworks, the study shows how complex evidence can be translated into clearer design reasoning that reflects operational constraints rather than generic wellbeing assumptions. The resulting PoV emphasises that effective EPSP wellbeing systems must minimise cognitive effort, adapt to unpredictable work contexts, and prioritise psychological safety and trust. Rather than relying on structured programmes, interventions are more effective when designed as low-friction, context-aware infrastructure that supports flexible and engagement. The UXR Play Cards developed through this work provide a reusable resource linking behavioural insights with practical implementation strategies.

Future work will focus on three areas: empirical validation through participatory workshops with EPSP stakeholders; comparison across multiple GenAI systems to examine interpretive differences and bias; and application of the AI-augmented UXR PoV approach to other high-stress domains to test transferability.

Overall, this work positions GenAI as a capability amplifier that supports analysis while preserving human responsibility for interpretation and design decision-making. Continued refinement aims to support more grounded and operationally realistic digital wellbeing design practices